\documentclass{article}

\usepackage{arxiv} 
\usepackage[utf8]{inputenc} 
\usepackage[T1]{fontenc}    
\usepackage{hyperref}       
\usepackage{url}            
\usepackage{booktabs}       
\usepackage{amsfonts}       
\usepackage{nicefrac}       
\usepackage{microtype}      
\usepackage{lipsum}		
\usepackage{graphicx}
\usepackage[super]{nth}
\usepackage[title]{appendix} 
\usepackage{chngcntr}  
\usepackage{xcolor}  
\hypersetup{
	colorlinks,
	linkcolor={red!50!black},
	citecolor={blue!50!black},
	urlcolor={blue!80!black}
}

\title{Mobility patterns of the Portuguese population during the COVID-19 pandemic}

\author{ \href{https://orcid.org/0000-0003-0502-6472}{\includegraphics[scale=0.06]{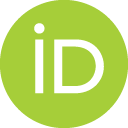}\hspace{1mm}Tiago~Tamagusko}\\
	Department of Civil Engineering\\
	University of Coimbra\\
	Coimbra, Portugal\\
	\And
	\href{https://orcid.org/0000-0002-1681-0759}{\includegraphics[scale=0.06]{orcid.png}\hspace{1mm}Adelino Ferreira}\thanks{adelino@dec.uc.pt} \\
	Research Center for Territory, Transports and Environment\\
	Department of Civil Engineering\\
	University of Coimbra\\
	Coimbra, Portugal\\
}

\renewcommand{\headeright}{preprint}
\renewcommand{\undertitle}{A preprint submitted for proceedings of the 2021 International Conference on Information Technology \& Systems}

\hypersetup{
pdftitle={Mobility patterns of the Portuguese population during the COVID-19 pandemic},
pdfsubject={physics.soc-ph},
pdfauthor={Tiago~Tamagusko and Adelino~Ferreira},
pdfkeywords={COVID-19; Mobility~Patterns; Rt; Changepoint; Modeling.},
}

\begin{document}
\maketitle

\begin{abstract}
SARS-CoV-2 emerged in late 2019. Since then, it has spread to several countries, becoming classified as a pandemic. So far, there is no definitive treatment or vaccine, so the best solution is to prevent transmission between individuals through social distancing. However, it is difficult to measure the effectiveness of these distance measures. Therefore, this study uses data from Google COVID-19 Community Mobility Reports to try to understand the mobility patterns of the Portuguese population during the COVID-19 pandemic. In this study, the \textit{Rt} value was modeled for Portugal. Also, the changepoint was calculated for the population mobility patterns. Thus, the change in the mobility pattern was used to understand the impact of social distance measures on the dissemination of COVID-19. As a result, it can be stated that the initial \textit{Rt} value in Portugal was very close to 3, falling to values close to 1 after 25 days. Social isolation measures were adopted quickly. Furthermore, it was observed that public transport was avoided during the pandemic. Finally, until the emergence of a vaccine or an effective treatment, this is the new normal, and it must be understood that new patterns of mobility, social interaction, and hygiene must be adapted to this reality.
\end{abstract}


\section{Introduction}
At the end of 2019, the new Coronavirus (SARS-CoV-2) appeared in the province of Wuhan (China) \cite{Huang2020a}, causing a disease named COVID-19 \cite{WorldHealthOrganizationWHO2020}. As a measure to combat COVID-19, China adopted the lockdown of this province on January \nth{23} \cite{Li2020}. This disease spread rapidly to other countries, with the first cases reported in Europe in the second half of January \cite{Spiteri2020}. Concerning Portugal, the first confirmed case of COVID-19 was on March \nth{3}, 2020 \cite{Portugal2020}, since the Portuguese government has adopted a series of measures to control the pandemic. To date, there are no vaccines for this disease, so the best strategy to combat COVID-19 is to prevent its transmission through social distancing. However, this is not a simple task, since a large part of the social activities are based on contact people and mobility. In the specific case of the transmission of COVID-19, the ideal scenario would be to monitor people’s contacts. Initiatives in this direction have been developed, but they face some concerns related to privacy. Another possibility is to measure the likelihood of contacts; this approach can be made by measuring the concentration of people in certain places. Thus, the population’s mobility patterns may indicate the degree of adoption of measures for social distancing \cite{Chinazzi2020a}. Nevertheless, effectively monitoring population mobility is a difficult task for governments. Google recently released the global time-limited sharing of mobility data \cite{GoogleLLC.2020}. This report presents several statistics and aims to promote studies that can help combat COVID-19. Mobility data is divided into six categories: retail and recreation; grocery and pharmacy; parks; transit stations; workplace; and residential. The values presented are percentage changes to normal (baseline) mobility patterns.
Currently, most European countries face the challenge of reactivating their economies; this task is linked to the gradual re-opening of services, public communal areas, and public transport. However, it is still not fully understood how the population has adopted the lockdown measures. In this sense, this paper finds relationships between the mobility patterns, the social distancing measures adopted, and the spread of the disease in Portugal.

\section{Materials and Methods}
\label{sec:methods}
\subsection{Events}
During the COVID-19 crisis in Portugal, the government adopted several measures to mitigate the spread of the disease. The main measures are grouped in Table~\ref{tab:tab1}. Other measures were adopted, but these events were considered more relevant.
\begin{table}
	\caption{Main public policies to mitigate the spread of COVID-19 in Portuga\cite{Portugal2020,Portugal2020a}	\label{tab:tab1}}
	\centering
	\begin{tabular}{lp{10cm}ll}
		\toprule
		Intervention     & Description     & Date (Y-m-d)\\
		\midrule
		Public events & Gatherings with more than 100 people forbidden.  & 2020-03-12     \\
		Social distancing    & Capacity restrictions in bars and restaurants, closed night clubs, limiting people in closed spaces are recommended. & 2020-03-12      \\
		Schools and universities     & Schools and universities closed.       & 2020-03-14  \\
		Social distancing     & Decrease in capacity to 1/3 and maintenance of a minimum distance of 1~m (ideally 2~m) in public services.      & 2020-03-17 \\
		Self-isolating of ill people     & Isolation is mandatory for sick people or being monitored by health authorities.      & 2020-03-19  \\
		Lockdown start   & Official start of the lockdown in Portugal.      & 2020-03-22 \\
		Public gatherings     & Gatherings of more than five people prohibited (except for large families).      & 2020-04-02  \\
		Lockdown end    & Official end of the lockdown in Portugal.      & 2020-05-03  \\
		\bottomrule
	\end{tabular}
\end{table}

\subsection{Data}
To develop this study, we used mobility data \cite{GoogleLLC.2020} and the cases of COVID-19 in Portugal \cite{Dong2020a,Roser2020}. The mobility report, called Google COVID-19 Community Mobility Reports, is data collected from mobile devices to quantify the movement of people during the pandemic. These values are anonymous and are aggregated based on the algorithm developed by Google, and the artificial noise sample is added to ensure that no individual can be identified based on their location information \cite{Aktay2020}. The report shows how the population moves and how long they stay in different locations (Table~\ref{tab:tab2}).
\begin{table}
	\caption{Report categories \cite{GoogleLLC.2020}}	\label{tab:tab2}
	\centering
	\begin{tabular}{lp{12.5cm}l}
		\toprule
		Category     & Subcategories  \\
		\midrule
		Retail and recreation & Restaurants, cafes, shopping centers, theme parks, museums, libraries, and movie theaters. \\
		Grocery and pharmacy & Grocery markets, food warehouses, farmers markets, specialty food shops, drug stores, and pharmacies. \\
		Parks & National parks, public beaches, marinas, dog parks, plazas, and public gardens. \\
		Transit stations & Public transport hubs such as subway, bus, and train stations. \\
		Workplace & Places of work. \\
		Residential & Places of residence. \\
		\bottomrule
	\end{tabular}
\end{table}

\begin{figure}
	\centering
	\includegraphics[width=\linewidth]{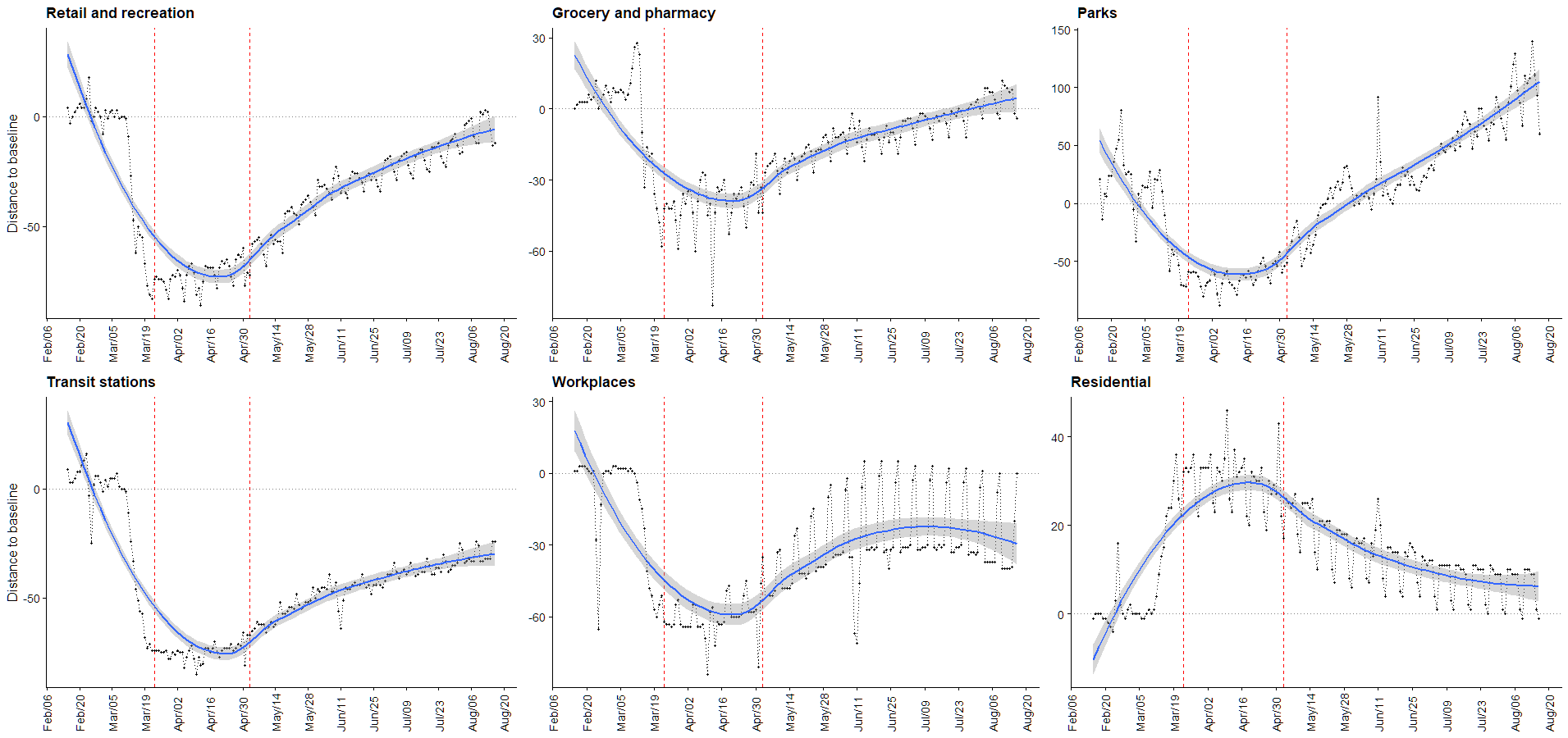}
	\caption{Mobility trends for Portugal.}
	\label{fig:fig1}
\end{figure}

The values presented for the categories are related to a baseline, which corresponds to the days of the week (from January \nth{3} to February \nth{6}, 2020). With these parameters, it is possible to assess the population’s adherence to the social isolation measures enacted by the government. The daily variation of values over time in Portugal, from February \nth{15} to August \nth{16}, is shown in Fig.~\ref{fig:fig1}.

\begin{figure}
	\centering
	\includegraphics[width=\linewidth]{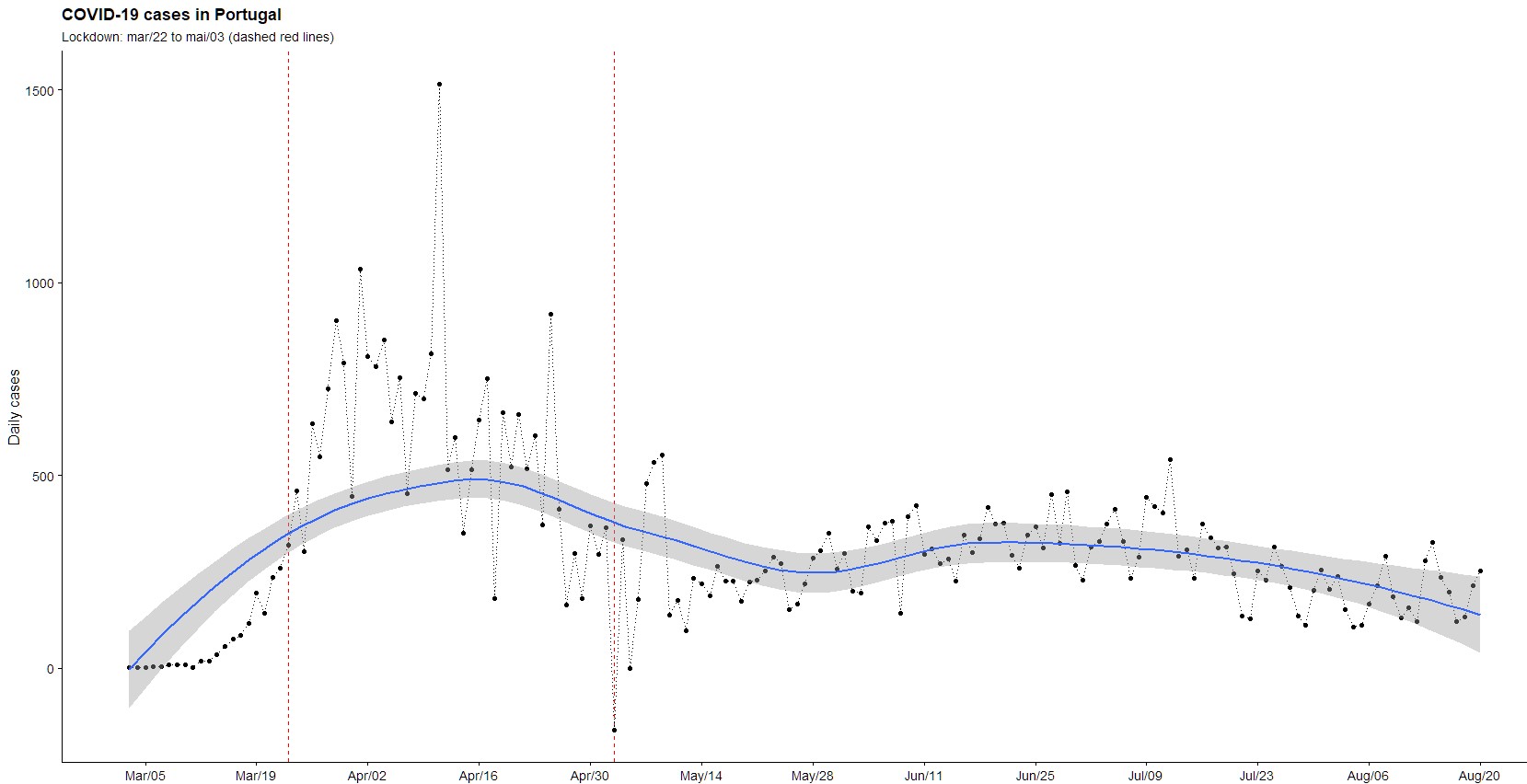}
	\caption{Cases of COVID-19 in Portugal.}
	\label{fig:fig2}
\end{figure}

In this graph, the vertical axis represents the distance to the baseline. Also, the area between red dashed lines represents the lockdown period in Portugal (started on March \nth{22} and finished on May \nth{3}). According to the data provided even before the lockdown, the values for the first five categories show falls. There was growth only in the residential category. In the days before the lockdown, there are peaks in the items Grocery and Pharmacy; this can be explained by the general rush to get supplies. The park-related peaks do not have a simple explanation. However, after this brief initial period, the population followed the imposed recommendations avoiding these locations during the lockdown. After the softening of the measures, there is an increasing demand (over 100\% at the end of August) for parks. It is assumed that influence of the adaptation of the population's routines to outdoor activities. It should also be noted that before March \nth{22}, schools and universities were closed, and several companies started to operate in teleworking. After lockdown, the values remained historically low. As expected, the tendency to stay at home is highly related to the workplace, in an approximately reversed trend.
In addition to the mobility data, values related to the cases of COVID-19 in Portugal were used between March \nth{3} (first confirmed case) and August \nth{20} (Fig.~\ref{fig:fig2}).

It is observed that the notification of the number of new cases is somewhat irregular. On weekends and holidays, the notifications are lower, and the following notifications are “inflated”. Another problem observed is that the number of confirmed cases is proportional to the number of tests performed. Therefore, the procedures adopted for testing the population influence the results of \textit{Rt} in this study.

\subsection{Methods}
The idea of this study is that \textit{Rt} can be influenced by the number of contacts between infected and susceptible individuals. Therefore, the social distancing measures adopted by the public authorities can influence this risk factor. In this study, the contact rate is approximated by the population’s mobility patterns during the pandemic period. Thus, it is considered that if the population decreases its presence in parks, restaurants, transportation stations, among others, the number of contacts decreases.
Supported by the \textit{R programming language} \cite{RCoreTeam.2017,RStudioTeam.2020}, the \textit{Changepoint} framework \cite{Killick2014} was used to detect changing values for mobility over time. Thus, it was possible to determine (approximately) the day when the mobility values changed their trend. Therefore, the goal is to detect the changepoint from the time series of mobility data provided. In this study, the mean approach was used, which uses the \textit{AMOC} (at most one change) method \cite{Hinkley1970} by default to detect a changepoints from the mobility patterns sample.
Another objective of this study is to calculate the \textit{Rt} in Portugal, a task that was developed with support from the \textit{R}, and from the \textit{EpiEstim} framework \cite{Cori2013}. \textit{Rt} is considered as to be the average number of secondary cases that each infected individual would infect if conditions remained as they were at time \textit{t}. Thus, the value of \textit{Rt} is determined according to Equation~\ref{eq:eq1} \cite{Cori2013}.

\begin{equation}
R(t)=\displaystyle\sum\limits_{s=1}^t I_{t-s} ws
\label{eq:eq1}
\end{equation}

Where \textit{I} is the number of people infected at any given time, and \textit{ws} corresponds to the probability of distribution of infections. This distribution depends on the characteristics of the disease. So, to determine the \textit{ws}, the method adopted was \textit{Non-Parametric SI}. We used the serial interval (SI) parameters presented by Nishiura et al. \cite{Nishiura2020}, with $\mu$ = 4.6 days (median serial interval) and $\sigma$ = 2.9 days (standard deviation).
Therefore, according to the study mentioned before \cite{Nishiura2020}, the average time for infected people to generate a second infection is 4.6 days. However, COVID-19 does not present itself equally in all infected individuals, they can be infectious over a period (serial interval). Consequently, it is expected that an individual exposed to COVID-19 may be infected and have an infectious window that lasts up to 14 days. This is the concept that endorses the WHO’s 14-day quarantine recommendation \cite{WHO2020}.
Finally, \textit{Rt} is an important indicator, as it can identify the stage of an infectious disease. For example, an \textit{Rt} of 2 means that each infected person, on average, transmits the disease to two other people. On the other hand, an \textit{Rt} less than 1 indicates that the spread of the disease is controlled, and it tends to disappear \cite{Wallinga2004}.

\section{Results}
The first result to be presented is the estimated day for a change in the behavior of the Portuguese population (Table~\ref{tab:tab3}). As presented before, these values were calculated using the \textit{Changepoint} framework \cite{Killick2014} and based on the daily values for mobility data in Portugal \cite{GoogleLLC.2020}.

\begin{table}
	\caption{Changepoint for mobility categories in Portugal\protect\footnotemark[\value{footnote}]}	\label{tab:tab3}
	\centering
	\begin{tabular}{ll}
		\toprule
		Category     & Changepoint (Y-m-d)  \\
		\midrule
		Retail and recreation & 2020-03-12 \\
		Grocery and pharmacy & 2020-03-13 \\
		Parks & 2020-05-20 \\
		Transit stations & 2020-03-13 \\
		Workplace & 2020-03-14 \\
		Residential & 2020-03-13 \\
		\bottomrule
	\end{tabular}
\end{table}
\footnotetext{Graphical representation in the appendix~\ref{sec:appendixB}.}

\begin{figure}
	\centering
	\includegraphics[width=\linewidth]{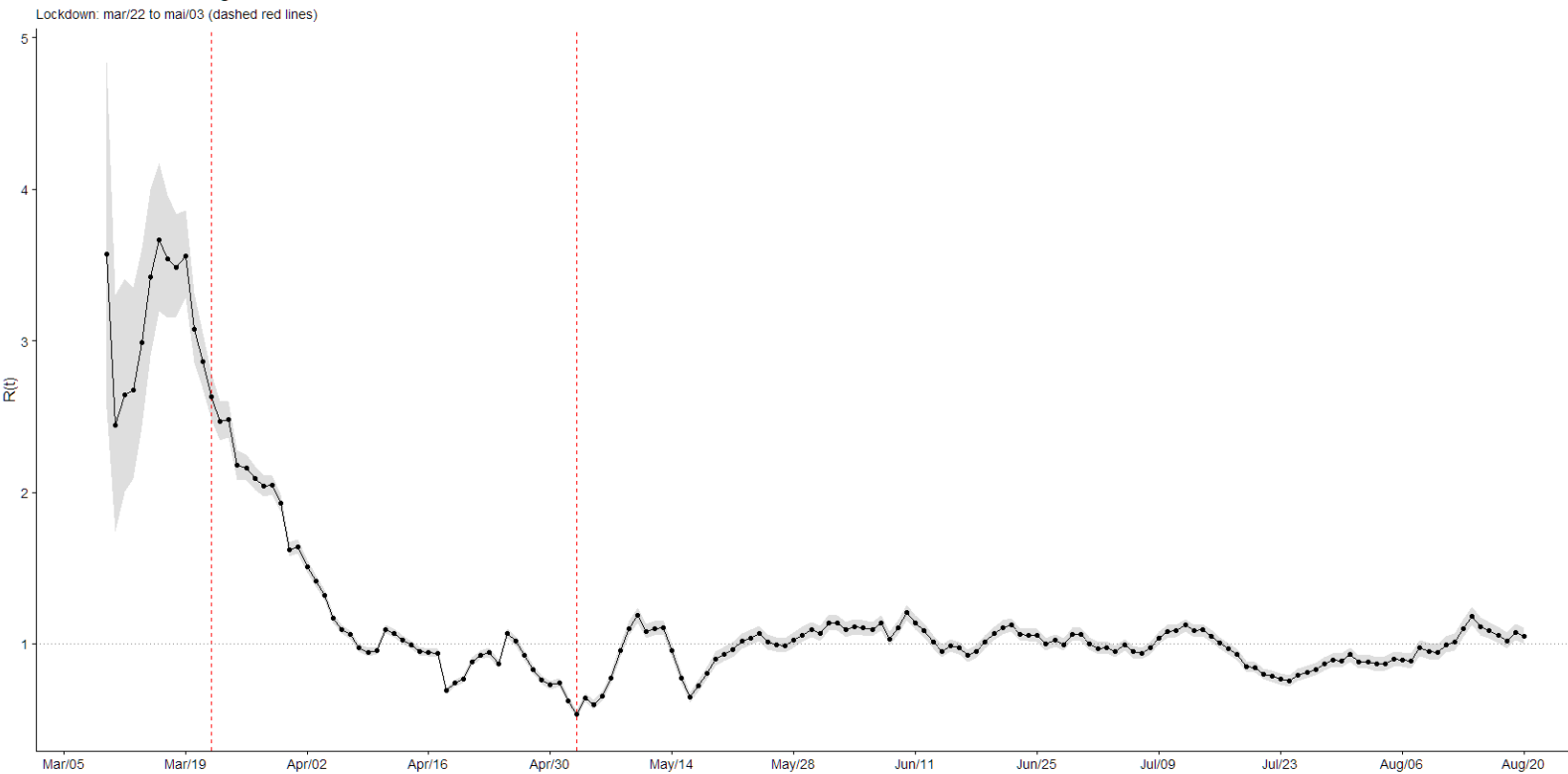}
	\caption{Estimated \textit{Rt} for Portugal}
	\label{fig:fig3}
\end{figure}

The approximate changepoint is between the \nth{12} and the \nth{14} of March. Still, it is observed that the most significant drop occurs from March \nth{12}, which is the date when the first public measures of social distancing were adopted. Thus, from the moment the government recommended people to stay at home, avoid public places, and maintain social distance, the population’s mobility pattern fell rapidly. Also, due to the characteristics of COVID-19, the number of new cases took time to slow down.
The first case of Covid-19 in Portugal was on March \nth{3}, however, the first calculated \textit{Rt} value is for March \nth{10}. Therefore, these first seven days are used by \textit{EpiEstim} to calculate $R_{0}$. Another observation is that the 95\% confidence interval (a grey area in the graph) is wide at the beginning of the pandemic in Portugal. The result of the \textit{Rt} value, based on epidemiological modeling\footnote{Result of the modeling developed in the appendix~\ref{sec:appendixC}.} developed for the COVID-19 pandemic in Portugal, calculated between March \nth{10} and \nth{20} August, is shown in Fig.~\ref{fig:fig3}.

It is possible to see that the measures adopted between March and April managed to bring the \textit{Rt} value from a level close to 3 for values below 1. However, after the end of the lockdown (May \nth{3}), the values have been close to 1, which means that the COVID-19 pandemic in Portugal has not yet been overcome. Likewise, in other European countries, measures of social distancing and lockdown were adopted, with similar \textit{Rt} results or even lower than those observed in Portugal \cite{Flaxman2020}.

\section{Discussion}
Considering a \textit{Rt} less than 1 to be an indicator of “control” of the pandemic, it can be seen that this value was reached only on April \nth{8} in Portugal, i.e., 25 days after the consolidation of the change in the behavior of the mobility patterns of the Portuguese population. Even after this date, the \textit{Rt} value was very close to 1, yet in some moments, it was above this threshold.
Also, the change in the population’s behavior (changepoint) happened before the lockdown. Therefore, this indicator may point out that even without the end of normal activities, people’s mobility is altered to adapt to the existing pandemic situation. Still, the places with the lowest flow of people during the monitored period are the transport stations. Nevertheless, this behavior of avoiding public transport creates a challenge for cities at this moment in the resumption of the economy.
Likewise, it must be understood that COVID-19 has not been eradicated in Portugal, and the second wave of contagions remains on the radar in Europe \cite{Wise2020}. Based on the values in Portugal, there is such a possibility since the number of confirmed cases so far is not likely to protect a population with herd immunity. Currently, it is essential to define red lines for the number of new daily cases. Similarly, successful measures used in other countries must be adopted.
Another critical point is that the \textit{Rt} value was obtained based on the number of infected individuals confirmed daily. These numbers may not correspond to the reality of the disease, because the number of confirmed infected depends on the number of tests performed, and the criteria adopted to test the population was not well explained.
\section{Conclusions}
As the main result of this study, it was observed that the Portuguese population reacted quickly, adopting social distancing, and changing their mobility pattern, even be-fore the government decreed restrictive measures. Still, it took 25 days for a \textit{Rt} value close to 3 to reach values near to 1. Now, it is expected that after the first wave of COVID-19, countries are better prepared for a probable second wave. Notwithstanding, observing the behavior adopted by the Portuguese population during that first lockdown, a second intervention of this type to be effective should last between two to four weeks.
It was also possible to observe that the sharpest drop occurred in public transport stations. Probably for fear of crowded locations, people sought individualized alternatives. A significant part of the population most likely used the car on their travels. With the re-opening of cities and the economy, this alternative may quickly prove unfeasible. Therefore, there is now a small window to co-opt users for active transport. Another observation was the significant increase in mobility in parks after the softening of lockdown measures. This trend of outdoor activities shows the importance of these spaces for cities.
Finally, we must understand that, for now, life cannot be as it was before the pandemic. Hence, until the discovery of a vaccine, the population, and the governments must be prepared for this new normal.

\begin{appendices}
\counterwithin{figure}{section}
\section{Data and reproducibility}
\label{sec:appendixA}
The raw data used for the development of this study are \cite{GoogleLLC.2020,Dong2020a,Roser2020}, the processed data and generated images can be accessed at \href{http://github.com/tamagusko/icts21}{github.com/tamagusko/icts21}. The codes developed in \textit{R} are available by request.
\section{Graphical representation of the changepoint}
\label{sec:appendixB}
\begin{figure}
	\centering
	\includegraphics[width=\linewidth]{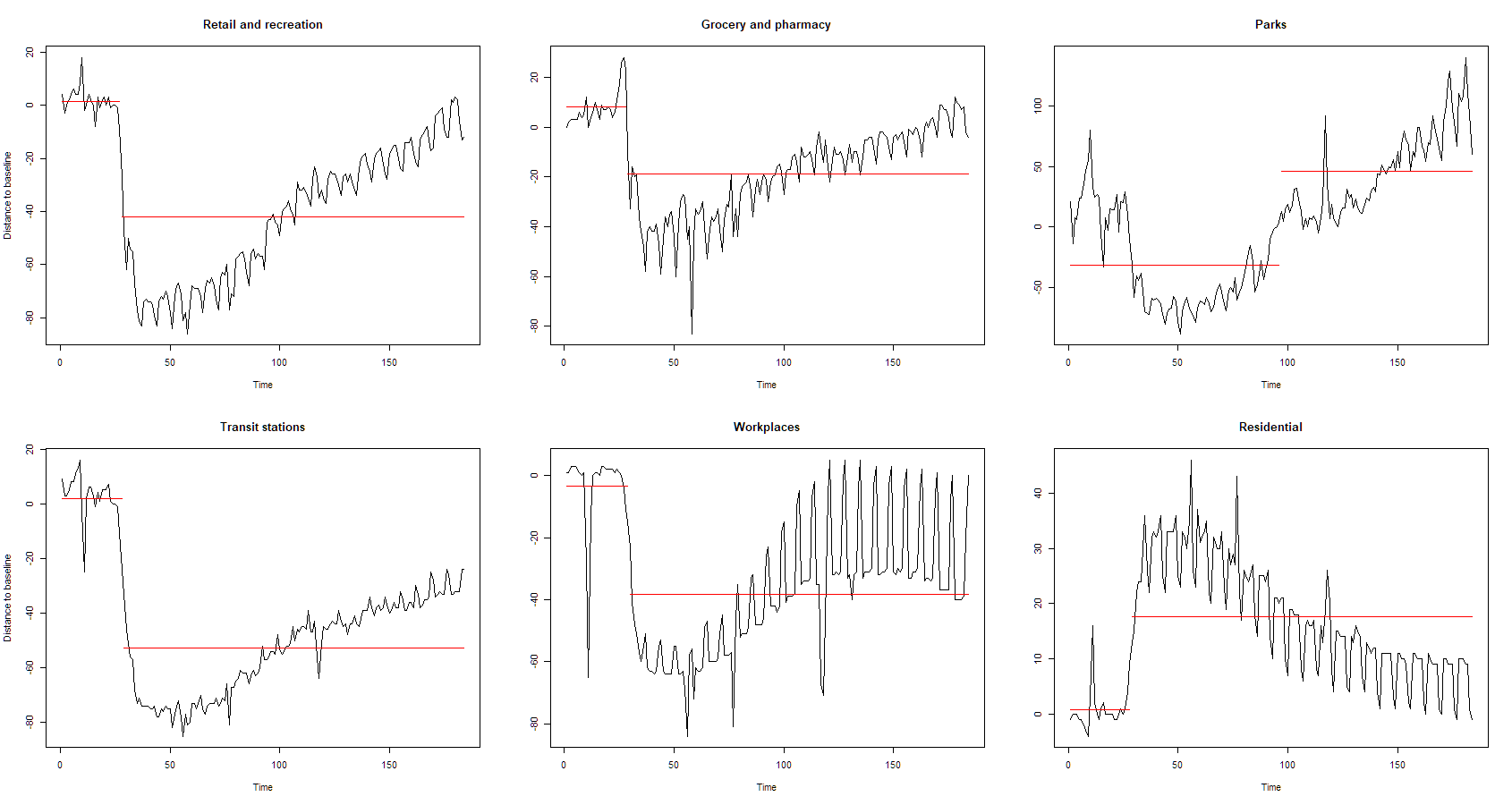}
	\caption{Changepoint in mobility trends for Portugal}
	\label{fig:figb1}
\end{figure}

These graphs (Fig.~\ref{fig:figb1}) support the interpretation of the data presented in Table~\ref{tab:tab3}. The approximate changepoint is on March \nth{13} (day 28) for most categories (excluding parks). Still, it is observed that the most significant drop occurs since March \nth{12}, this being the day of the first measures of social distance in Portugal.

\section{Result of epidemiological modeling}
\label{sec:appendixC}
\begin{figure}
	\centering
	\includegraphics[width=\linewidth]{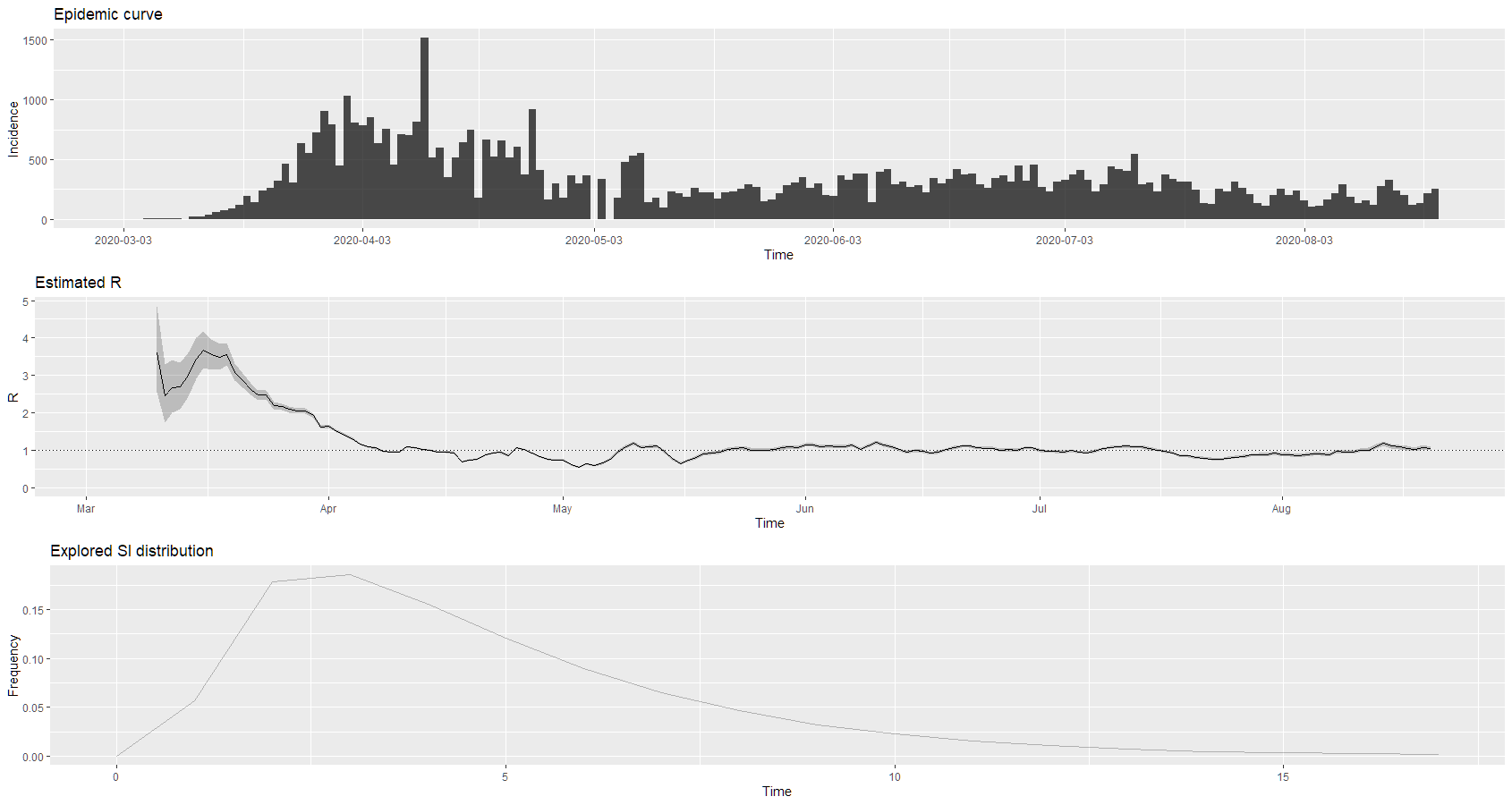}
	\caption{Result of epidemiological modeling}
	\label{fig:figc1}
\end{figure}

Epidemic curve, estimated \textit{Rt} value, and serial distribution (Fig.~\ref{fig:figc1}) developed with support from the \cite{Cori2013}.

\end{appendices}

\end{document}